\documentclass[pra,aps,superscriptaddress,groupedaddress, twocolumn]{revtex4}

\usepackage{mathrsfs} 
\usepackage{xspace,amsmath,amsfonts,amsthm,amssymb,amsbsy}
\usepackage{graphicx}
\usepackage{amssymb}
\usepackage{color}
\usepackage{psfrag}
\usepackage{ifsym}
\usepackage{epstopdf}
\usepackage{units} 
\usepackage{bbm}
\usepackage{bbold}
\usepackage[usenames,dvipsnames]{xcolor}
\usepackage[colorlinks=true,citecolor=Cerulean,linkcolor=RubineRed,urlcolor=Cerulean]{hyperref}

\graphicspath{{../images/}}

\begin{document}

\author{Aur\'elia Chenu}
\email[]{achenu@mit.edu}
\affiliation{Massachusetts Institute of Technology,  77 Massachusetts Avenue, Cambridge, MA 02139, USA}
\affiliation{Singapore-MIT Alliance for
Research and Technology, 138602 Singapore}

\author{Nir Keren}
\author{Yossi Paltiel}
\affiliation{Department of Plant and Environmental Sciences, Alexander Silberman Institute of Life Sciences, Givat Ram, The Hebrew University of Jerusalem,  91904 Jerusalem, Israel}

\author{Reinat Nevo}
\author{Ziv Reich}
\affiliation{Department of Biomolecular Sciences, Weizmann Institute of Science, 7610001 Rehovot, Israel}

\author{Jianshu Cao}
\email[]{jianshu@mit.edu}
\affiliation{Massachusetts Institute of Technology,  77 Massachusetts Avenue, Cambridge, MA 02139, USA}
\affiliation{Singapore-MIT Alliance for Research and Technology,  138602 Singapore}

\title{Light Adaptation in Phycobilisome antennas: \\
Influence on the Rod Length and Structural Arrangement}


\begin{abstract}
Phycobilisomes, the light-harvesting antennas of cyanobacteria, can adapt to a wide range of environments thanks to  a composition and function response to stress conditions. We study how structural changes influence excitation transfer in these super-complexes. Specifically, we show the influence of the rod length on the photon absorption and subsequent excitation transport to the core. Despite the fact that the efficiency of individual disks on the rod decreases with increasing rod length, we find an optimal length for which the average rod efficiency is maximal. Combining this study with  experimental structural measurements, we propose  models for the arrangement of the phycobiliproteins inside the thylakoid membranes, evaluate the importance of rod length, and predict the corresponding transport properties for different cyanobacterial species. This analysis, which links the functional and structural properties of full  phycobilisome complexes,  thus provides further rationals to help resolving their exact structure. 
\end{abstract}

\maketitle

Photosynthetic organisms have evolved efficient strategies to harvest sunlight, directing the excitation through various molecular aggregates to the reaction center, where the light energy is converted into  electrical and chemical energy. 
Among the primary producers, cyanobacteria, which  can be viewed as the most important group of organisms ever to appear on our planet \cite{Herrero2008a}, show great versatility in maintaining their structures. 
  The cyanobacterial light-harvesting antenna, phycobilisome, is adapted to the particular environmental  conditions, with drastic composition and function changes under stress conditions \cite{Raps1985a, Grossman1993a, Bar-Eyal2015a}. 
Its high performance under drastically different living conditions has spurred investigations on the  excitation transfer properties of this super-complex.

  \begin{figure}[t]
 \includegraphics[width=0.8\columnwidth]{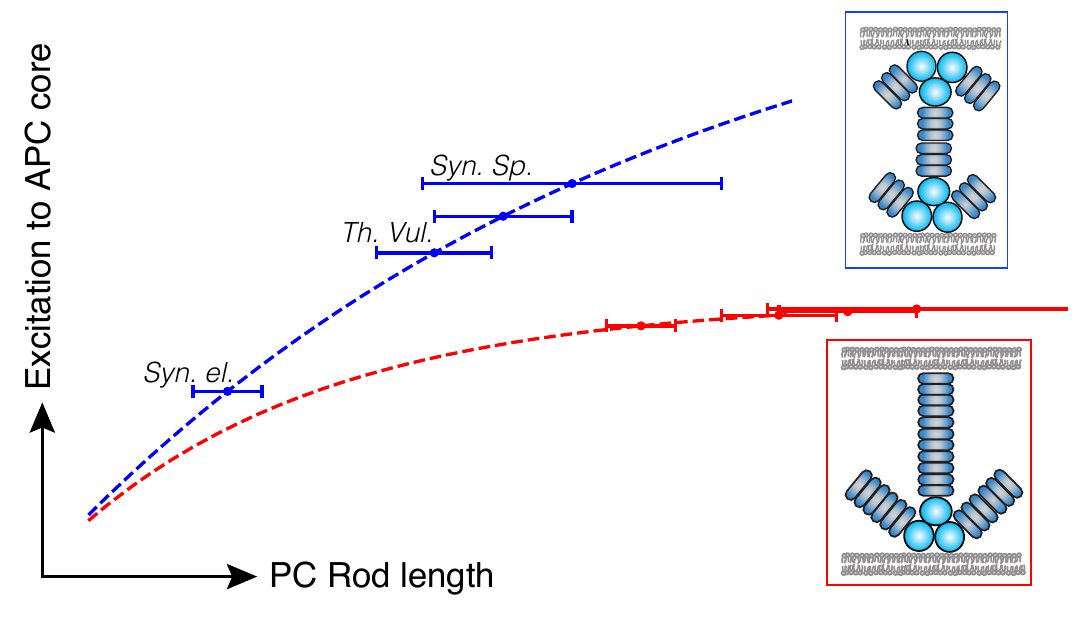}
 \caption{Graphical abstract illustrating the different structural arrangements of PBS in-between the thylakoid membrane  proposed here, and the corresponding kinetic analysis. }
 \end{figure}

The performance of  light-harvesting antennas to transfer the electronic excitation to the reaction center is related to the spatial arrangement of the chromophores and proteins forming them, and many works have studied the transport properties in natural networks for a variety of organisms \cite{Falkowski1981a, Cogdell2008a, Plenio2008a, Scholes2012a, Moix2011a,Wu2012a, Duffy2013a, BlankenshipBook}. Considering that increasing the area occupied by the chromophores increases the cross-section to capture photons but decreases the transfer efficiency to the reaction center, an optimal efficiency is expected to be  established for a particular ratio between donors (chromophores of the antenna) and acceptors (chromophores of the reaction center). 
Such observations have inspired the design of artificial networks \cite{Walschaers2013a} and light-harvesting devices \cite{Kim2001a, Polivka2004, Balaban2005a},  with promising applications in the development of organic solar cells \cite{Bredas2009a, McConnell2010a, Blankenship2011a} and nano electronics \cite{Salleo2015a,Rivnay2014a}. 

Phycobilisomes (PBS) are mostly composed by the assembly of phycocyanin (PC), and sometimes phycoerythrin (PE), hexamers into rods, linked to a core formed by allophycocyanin (APC) trimers \cite{Adir2008a}.
The environmental factor that most influences the structure of PBS is light intensity, which can change the PC:APC ratio, and yield a maximum production of PC under optimal photon flux \cite{deLorimier1992a}. 
Other environment factors, such as light color and temperature, also influence the composition of the rods, changing the PC:PE ratio, either as an adaptation to the incident spectra \cite{Stowe2011a,Wiltbank2016a,Kehoe2006a, Wiltbank2016a}, or as a response to perturbed metabolic processes that regulates the synthesis of phycobiliproteins.

 Here, we study the adaptation of the PC:APC ratio as a response to the light intensity, and, combining with experimental measurements, we propose structural models for the arrangement of the PBS complexes in between the thylakoid membranes. To do so, we study the influence of the number of hexamers forming an homogeneous PC rod on the efficiency of excitation transport to the APC core. We first briefly describe the structure of phycobilisomes and their assembly properties. We then present  the model and results of transport in the full aggregation state. In particular, we show how the amount of excitation transferred to the  APC  core can be kept constant under decreasing light intensity,  increasing the PC rod length. We further provide different models for the structural arrangement of the PBS complexes, and compare the corresponding transport properties based on measurement for different cyanobacterial species. 


\section{Structure of phycobiliproteins}
The phycobilisomes \cite{Adir2008a, Chakdar2016a} 
consist of various types of bilin cofactors and pigment-proteins, known as phycobiliproteins (PBPs), classified into four major groups: allophycocyanin (APC $\lambda_{\rm max}$=652 nm), phycocyanin (PC  $\lambda_{\rm max}$=620 nm), phycoerythrin (PE $\lambda_{\rm max}$=560 nm) and phycoerythrocyanin (PEC $\lambda_{\rm max}$=575 nm). 
The phycobilin chromophores being linear tetrapyrroles, they are conformally highly flexible, with optical and photo-physical properties highly dependent on their environment. Hence, the pigment-protein interaction plays a crucial role in the function of PBP as antenna complex, 
such that the chromophores' absorption and emission spectra in isolated state or de-naturated PBP differ from the properties in intact form.

\begin{figure}
\includegraphics[width= 1\columnwidth]{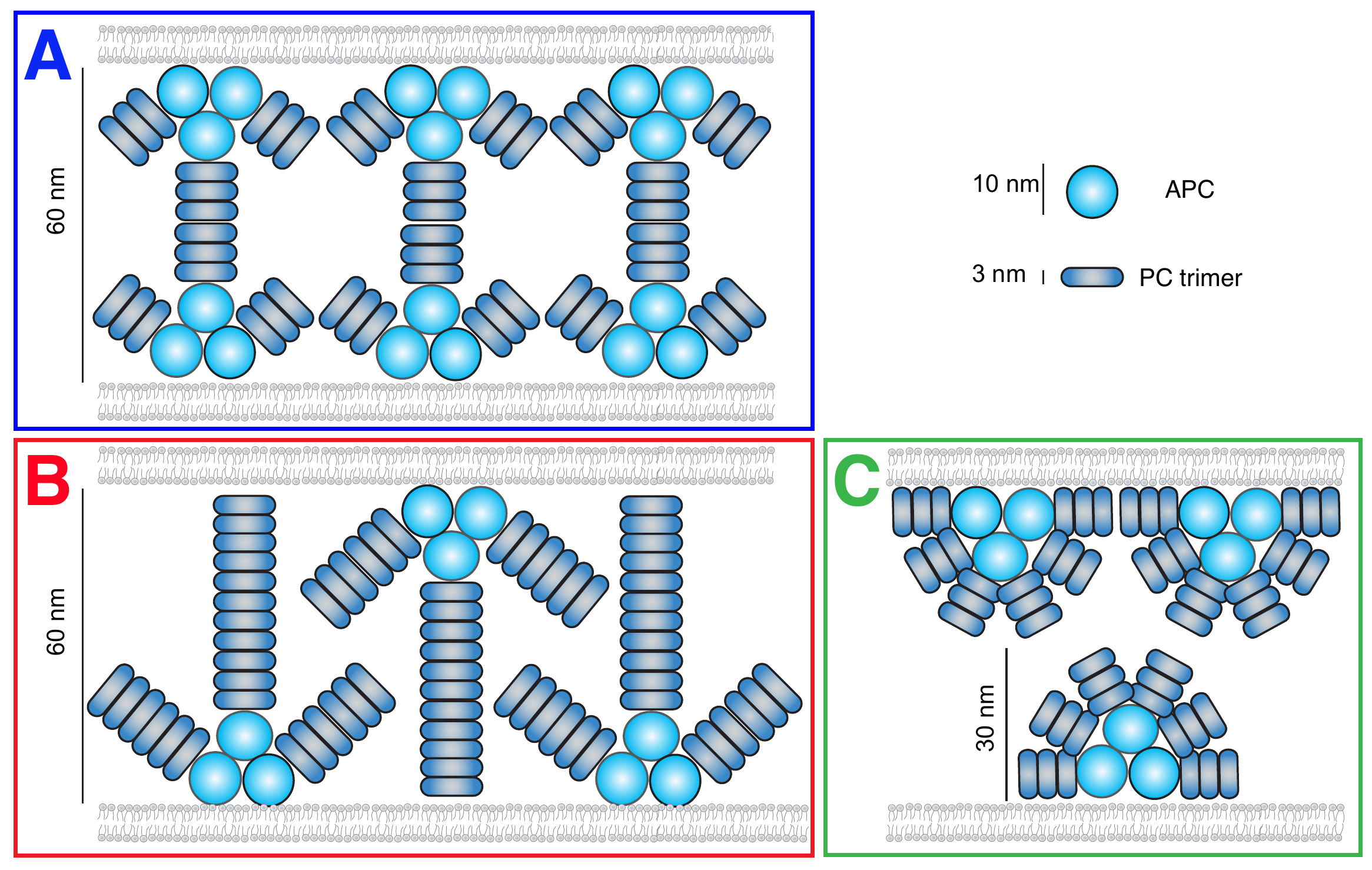}
\caption{Cartoon of the proposed two limiting stacking models (A and B) of PBS units in the space between thylakoid membranes, along with the currently used model (C, adapted from Ref. \cite{Olive1997a}). The length of the PC rod can be inferred from TEM images, and is given in Tab. \ref{tab} for different species.  \label{fig:stacking}}
\end{figure}

All PBPs are  formed from a heterodimer primary building block, known as the $(\alpha \beta)$ monomer,  composed of two homologous subunits  ($\alpha$ and $\beta$ helices) and bilin cofactors:
namely,  $\alpha_{84}$ and $\beta_{155}$ for APC, with an additional $\beta_{84}$ chromophore for PC. 
The high degree of homology between the subunits and the monomer allows for further self-assemblies \cite{Adir2006a}, which leads to the formation of complex structures with controllable and efficient properties. 
PC, PEC and PE associate into $(\alpha \beta)_6$ hexamers, and further stack into \emph{rods}.
 APC trimers  associate differently, forming cylinders from four trimers. These cylinders further pack into the \emph{core} of the PBS. An APC core and rods made of stacked disks of PC and PE proteins further form a phycobilisome (Fig. \ref{fig:stacking}),  which is attached to the stromal side of the thylakoid membrane through the APC proteins in its core.

Compared to the Chl- and BChl-based antennas, the PBSs are characterized by large inter-chromophore distances. For example, in the minimal monomeric subunit of \textit{Thermosynechococcus vulcanus}, the distance between the photo-sensitive bilins is 40\AA~   (between $\alpha -\beta$) or 50\AA~ ($\beta - \beta$) [see Fig. S1 in supplementary material (SM)]. While assembly into trimer or hexamer results in more packed structures  \cite{Adir2001a}, the distances ($\sim$20\AA) remain much larger that in the chlorophyll-based antenna (e.g. those of green sulfur bacteria or plants have a typical distance of 10\AA). 
Despite the large distances, the transfer rates remain extremely efficient, with an overall quantum yield above 95\%, which suggests that efficient transport does not require a high density of chromophores. While the energy transfer in PBS has been extensively studied on isolated subunits (e.g. see Refs. \cite{Holzwarth1987a, Suter1987a, Sauer1988a, Debreczeny1993a} and the rates in Table~S1, SM), it is still unclear how further aggregation affects the energy flow.  Very little is known of the excitation energy transfer at the level of rod aggregation, which we address here, in the framework of incoherent transfer.

\section{Excitation transfer in phycobilisome rods}
We calculate the transfer of excitation along a linear rod of PC terminated by an APC trap, and study the transfer efficiency as a function of the homogeneous rod length and light intensity. We show that, for a given light intensity, there exists a rod length optimizing  transfer to the core, and that the total excitation transferred to APC can be kept constant for different light intensities, by adjusting the rod length. The assumptions of our kinetic model are as follow: (i)  we use incoherent, F\"orster theory considering that the distances between chromophores are large; (ii) we exclude back-transfer from the core to the rod due to the small spectral overall of the components and the quick depopulation of APC \cite{Glazer1985a, Grabowski1978a}; (iii)  the spectral properties of each chromophores composing the PC trimers are assumed to be identical throughout the rod, which is reasonable for a given aggregation form. While this assumption neglects  static disorder along the rod, it does not affect our results qualitatively. The different forward and backward rates are obtained from the geometry and the resulting dipole-dipole interaction. 
Because there are little details of energy transfer within the APC core, we focus on the PBS rod. Our results can be extended to heterogeneous  rods adding phycoerythrin as building blocks.

\begin{figure}
\includegraphics[width=0.7\columnwidth]{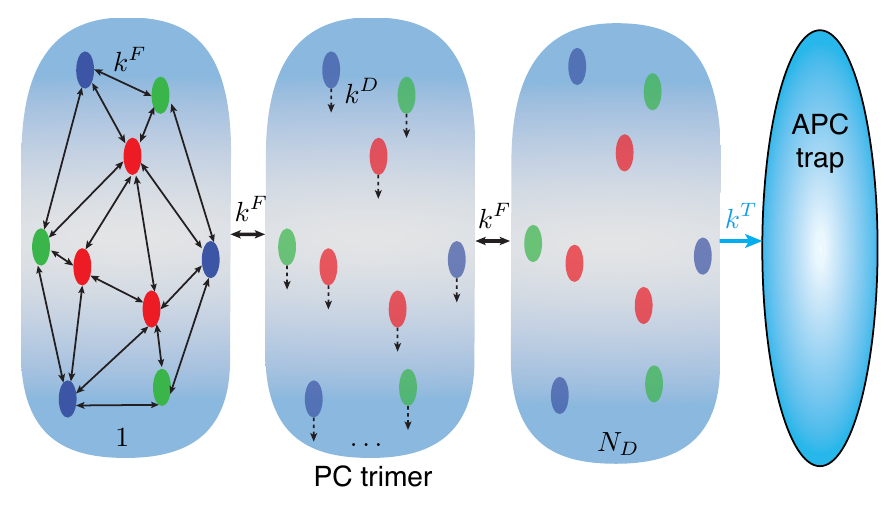}
\caption{Cartoon of the kinetics scheme modelled here representing the  rates for the chromophores ($\alpha_{84}$, green, $\beta_{84}$, red and $\beta_{155}$, blue) forming each PC trimer. All chromophores transfer via F\"orster rates ($k^F$), and have a decay rate $k^D$ accounting for quenching and spontaneous decay. Only the chromophores forming the last trimer, $N_D$, transfer to the APC trap, through $k^T$. 
\label{fig:kinetics} }
\end{figure}

The population dynamic along the rod is governed by the rate equation 
$\dot{P}(t) = - K P(t)$. 
The rate matrix, 
\begin{eqnarray}
K_{n\rightarrow m} &=& - k^F_{m\rightarrow n} + \delta_{nm}\big( k^D  + \sum_{m'} k^F_{n \rightarrow m'}+ \delta_{N, N_D }k^T \big) \nonumber  \\
&&-  \delta_{N, N_D } \delta_{n,{\rm APC}} k^T,
\end{eqnarray}
$n$ labeling the chromophores on the $N$-th PC trimer,  includes transfer rates to neighbouring units ($k^F$ obtained from F\"orster theory), decay rates ($k^D$) accounting for  quenching and radiative decays, and trapping rates ($k^T$) from the PC trimer terminating the rod, $N_D$, to the APC core (see Fig.~(\ref{fig:kinetics}) for a cartoon of the kinetic scheme). 
The energy flow between PC rod and APC core being still unresolved, partly because of complex structure that could involve linker proteins  \cite{Liu2005a, Tal2014a, David2014a}, we use a phenomenological trapping rate from the last PC disk to APC,  taking $k^T = 0.025$ps$^{-1}$, which  corresponds to the lowest range of the experimentally measured range, (0.025-0.056 ps$^{-1}$), \textit{cf.} \cite{Li2004a}.

Considering the large distances separating the chromophores, the PC-PC transfer rate are evaluated from F\"orster theory (see \cite{Foerster1965a} and, e.g.,  \cite{Scholes2003a}  for a review), which gives the rate between a donor $D$ and an acceptor $A$ as 
\begin{equation} \label{eq:rate}
k^F_{D\rightarrow A} = \frac{1}{\hbar^2 c} \left| V_{dd} \right|^2 \frac{\int_0^\infty d\omega E^D(\omega) I^A(\omega)}{\int_0^\infty E^D(\omega) d\omega \int_0^\infty I^A(\omega) d\omega}.
\end{equation}
 The dipole-dipole coupling, $V_{dd}= \frac{\mu_D \mu_A}{4 \pi \epsilon_0 n^2}\frac{\kappa_{DA}}{R^3_{DA}}$,  in units of cm$^{-1}$, and the orientation factor, $\kappa_{DA} = \hat{\mu}_D \cdot \hat{\mu}_A - 3(\hat{\mu}_D \cdot \hat{R}_{DA} ) (\hat{\mu}_A \cdot \hat{R}_{DA} )$, are  obtained from the structural data of the  cyanobacterium {\it Thermosynechococcus vulcanus} \cite{Adir2001b} (also see Fig. S1 in SM).  The orientations of the  transition dipole moments were estimated by fitting a line through the conjugated portions of the bilin chromophores. 
 In addition to the symmetry used to build the hexamer, the rod is constructed repeating the hexamer disks with a pitch of 60.5\AA   \
  along the rod direction \cite{Schirmer1986a}.  
 We use the screening factor and calculated dipole strength from Ref. \cite{Ren2013a}, i.e. $n^2 =2$ and $\mu= 13 $D for all chromophores --- note that lower values  have sometimes been used in the literature \cite{Womick2009a}. Quenching and radiative decay are taken from the experimental fitting of Ref. \cite{Eisenberg2017a}  as a biexponential decay, with time constants of 0.8ns and 1.7ns, and a relative weight of 60\% and 40\%, respectively.

As mentioned above, the spectral properties of the bilin pigments highly depend on their environment, i.e. on the cyanobacteria strain and the aggregation form. Thanks to a recent theoretical analysis of experimentally spectra in trimeric and hexameric forms \cite{Eisenberg2017a}, the chromophore spectra in aggregate conformation are now available for \textit{Th. vulc.} cyanobacterium.  In the following, we will use these spectra---LS solvent, wet phase. 
We note that they are significantly different from  those measured for isolated chromophores \cite{Debreczeny1993a}. This difference is likely due to denaturation of the protein environment following the isolation process.

The detailed rates between each chromophore of the PC rod are given in the SM. The results are  in quite good agreement with rates calculated for different species up to the hexameric structure, e.g. \cite{Sauer1988a}, especially 
 considering that the rates are extremely sensitive to the structure. 
 This is best illustrated in Ref. \cite{Sauer1988a}, where the rates for  \textit{A. quadruplicatum} are given for monomer, trimer and hexamer aggregation states, for both the `old' \cite{Schirmer1986a} and the refined structure \cite{Schirmer1987a}.  We see a variation of more than two-order of magnitude variation in the monomer rate, and a 25-fold factor in some of the trimer and hexamer rates for the different structures, although they correspond to the same species. The rate are therefore highly dependent on the structure resolution, and all the more so on the species.

To get the transfer efficiency, we first consider the case where the excitation is harvested by the $i$-th trimer disk, and solve the dynamics  for any rod length. 
Considering incoherent transport, the trap population collected in APC at steady state is obtained from  
$\int_0^\infty \dot{P}_{\rm APC}(t) dt= \delta_{N,N_D} \sum_{n=1}^{9}  k_n^{T} P_n(t)$. 
 It can be integrated directly using the rate equation to evaluate the population. This  yields to  the disk efficiency, defined as the   efficiency of transfer after excitation of the $i$-th disk, as 
\begin{equation} \label{eq:etai}
\eta^{(i)} =  \delta_{N,N_D} \sum_n k_n^{T} \left( K^{-1} . P^{(i)}(0)\right)_n, 
\end{equation}
where $P^{(i)}(0)$ corresponds to an initial normalized excitation of disk $i$. 
The  efficiency of the rod is then given by the average efficiency weighted by the harvesting cross-section  \cite{Kim2010a}, which is $N_D / (N_D + 1)$ in a rod of $N_D$ PC trimers and one APC core. This gives the rod efficiency as    
$\eta = \frac{1}{N_D+1 } \sum_{i=1}^{N_D} \eta^{(i)}$. 

\begin{figure}
\includegraphics[width=1\columnwidth]{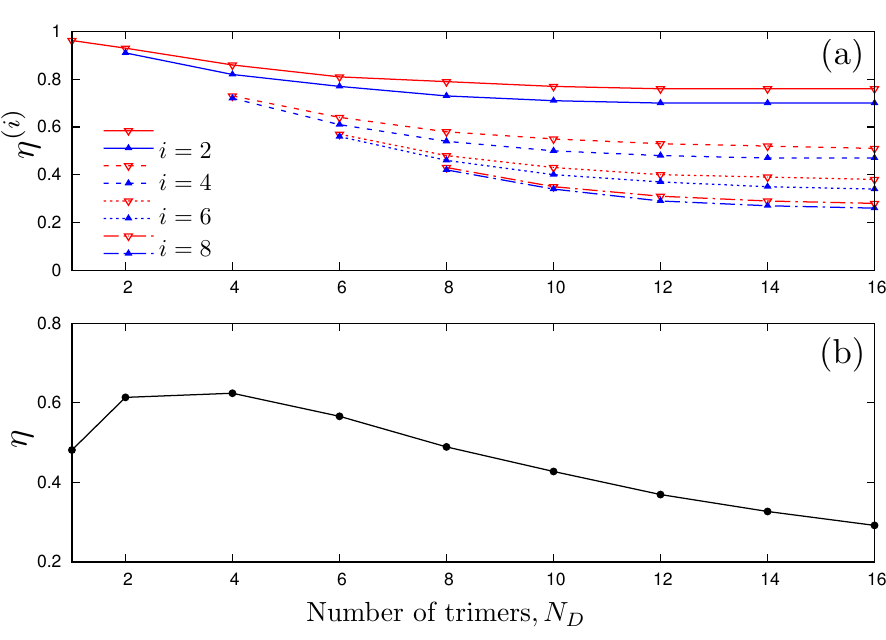}
\caption{Efficiency (a) of the PC-trimer disk, $\eta^{(i)}$ [Eq. (\ref{eq:etai})], for different locations of the initial excitation and (b) of the rod, $\eta$, defined as the efficiency of harvesting and transferring one excitation to the core, as a function of the number of PC trimers.   Despite a disk-efficiency decreasing with the rod length, the rod-efficiency displays a maximum for a length of $\sim$2-4 trimers due to an increased harvesting cross-section. 
\label{fig:efficiency} }
\end{figure}

The efficiency of transferring one excitation as function of the rod length,  defined  in Eq. (\ref{eq:etai}), is shown in Figure (\ref{fig:efficiency}a).  It decreases as the rod length increases, even if the excitation is captured in the trimer neighbouring the APC trap (see e.g. $\eta^{(1)}$ in Fig. \ref{fig:efficiency}a). This is due to leakage of the excitation to the neighbouring trimer,  opposite to the trap. 
As the excitation is captured further away from the trap, the transfer efficiency decreases due to losses in the form of quenching to the protein and radiative decays. 
The rod averaged efficiency, $\eta$, however, also accounts for the absorbing power that increases with the rod length. 
Because of a trade-off between the increasing absorbing power and the decreasing transport efficiency, there exists an optimal rod length for which the efficiency is maximal. 
 For the set of chosen parameters, we find this optimum is reached with $N_D$=4 trimers---see Fig. (\ref{fig:efficiency}b).  Results of the harvesting time, presented in the SM, further support this feature. 
Note that the results do not change significantly using a trapping rate of 0.056 ps$^{-1}$. 

\begin{figure}
\includegraphics[width=1\columnwidth]{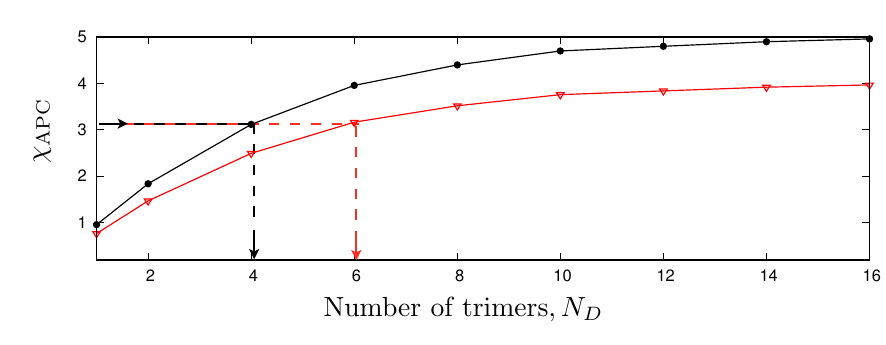}
\caption{Total excitation transferred to the APC trap at steady-state for 100\% (black) and 80\% (red) light intensity. The arrows indicate that, to keep a constant excitation reaching APC under varying excitation intensity, the rod length must be increased for decreasing light intensity (e.g. from 4 to 6 trimers in the particular example illustrated here).    \label{fig:pop} }
\end{figure}

The total excitation transferred to the APC trap is readily given by $\chi_{\rm APC}=\sum_i \eta^{(i)}$. Figure (\ref{fig:pop}) shows that the excitation of the APC trap increases as the rod length increases due to higher absorption cross-section. A lower intensity flux results in a lower transfer of excitation. It is possible to compensate for lower light intensity by increasing the rod length. As illustrated in  Fig. \ref{fig:pop}, maintaining the excitation of APC constant for 80\% intensity can be done increasing the rod length from 2 to 3 hexamers.  

Note that the results presented here are in line with that of Ref. \cite{Kim2010a}, now extended to a biological system. There, the optimal length for a donor-acceptor system had been found  in a 1D geometry. Here, we extended this to a quasi-2D geometry, and showed that there is an optimal number of donor disks for which the transfer to the acceptor is optimal. 
In addition, the APC population can be fitted into a bi-exponential (not presented here), such that two rates are mostly sufficient to account for the direct and indirect transfers, where the slower rate integrates all back transfers, which is in agreement with Refs. \cite{Holzwarth1987a, Suter1987a}.

\begin{table}[b]
\begin{tabular}{clccc}
 & Cyanobacterial species  &	Distance (nm) & $N_D^A$  & $N_D^B$ \\
\hline
(c) & Synechocystis sp. PCC6803 & 31$\pm$5 & - & -\\
&\hfill olive mutant	(APC only)& &&\\
(d) & Synechococcus elongatus PCC7942 &	45$\pm$3              & 4 &  10\\
(b) & Thermosynechococcus vulcanus	 & 53$\pm$5 & 7 & 12\\
(e) & Gloeobacter violaceus PCC7421 &	55$\pm$6     & 8 &13\\
(a) & Synechocystis sp. PCC6803	& 58$\pm$13        & 9 & 14 \\ 
\end{tabular}
\caption{Space between the thylakoid membranes measured from the cryo-fixed samples presented in Fig. \ref{fig:TEM}. The number of PC trimers for model A and B, $N_D^{A}$ and $N_D^{B}$, respectively, has been inferred from the intra-membrane width without the APC complexes---see text for details.  \label{tab}} 
\end{table}

\begin{figure*}
\includegraphics[width= 1\textwidth]{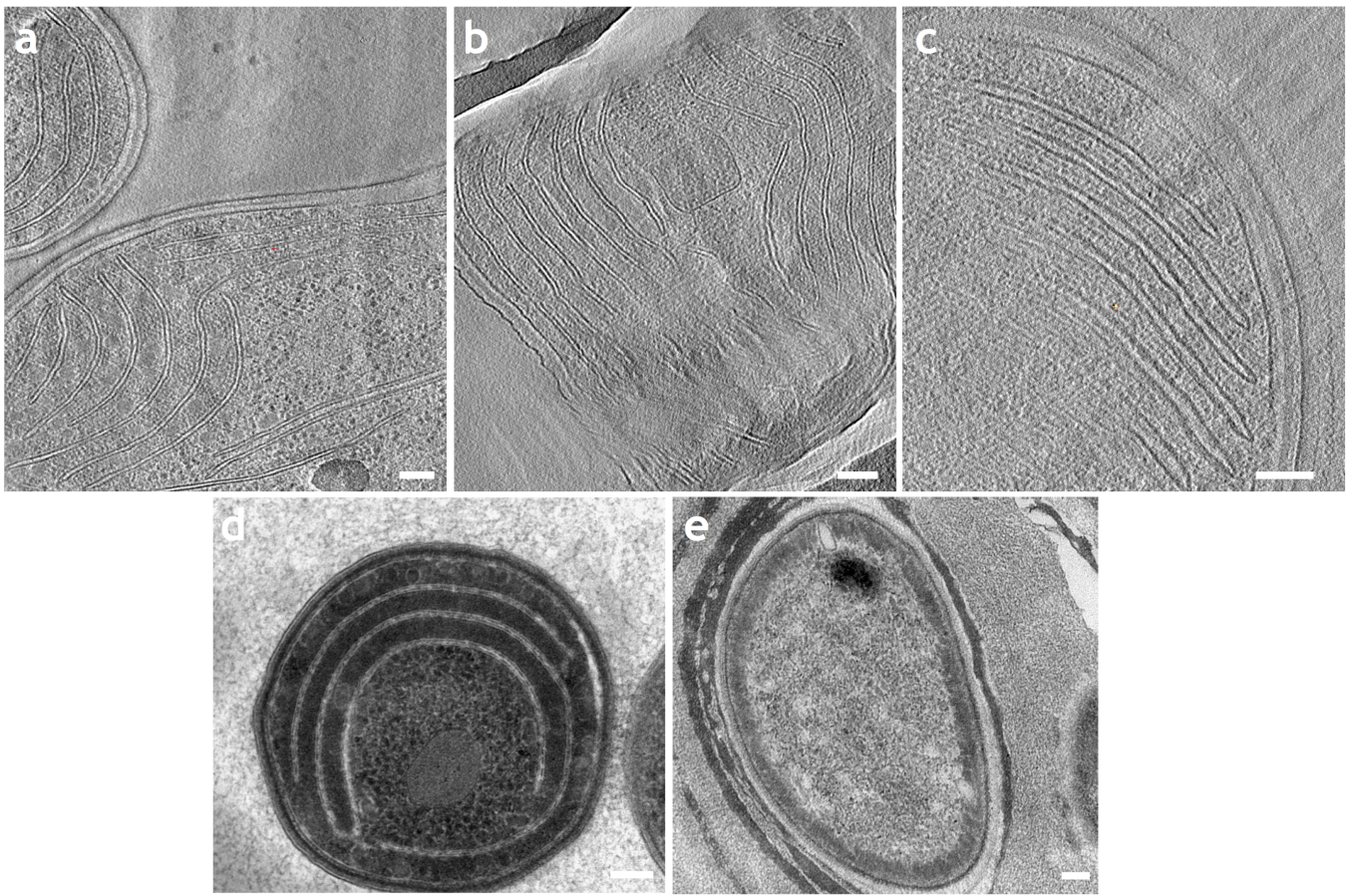}
\caption{Thylakoid membrane organization of the organisms described in Table \ref{tab}. (a) \textit{Synechocystis sp.} PCC6803; (b) \textit{Thermosynechococcus vulcanus}; (c) \textit{Synechocystis sp.} PCC6803 `olive' ($\Delta$PC) mutant, which is deficient in plastocyanin and, hence, lacks the peripheral rods of the phycobilisome \cite{Elmorjani1986a,Olive1997a}; (d) \textit{Synechococcus elongatus} PCC7942; (e) \textit{Gloeobacter violaceus} PCC7421. (a-d) Tomographic slices ($\sim$12 nm-thick) of vitrified (a, b, c) or cryo-immobilized, freeze-substituted (d) samples. (e) TEM image of cryo-immobilized, freeze-substituted cells. Bars: 100 nm.
\label{fig:TEM}}
\end{figure*}

\section{Structural considerations for the PBS arrangement \label{sec:discussion}} 
The exact structure of PBS complexes is still an open question. Complete PBS structures were not resolved by X-ray crystallography, despite intensive efforts \cite{Marx2014b}. The best evidence currently available are from TEM image reconstruction studies. 
Current resolutions, of up to 3 PC hexamers \cite{Arteni2009a}, and  2 PC hexamers \cite{Chang2015a}, still offer very limited structural models. We here use the transport properties to provide additional argument.

Considering that the space between the thylakoid membranes is most likely packed with phycobiliproteins, we can  infer the PC rod sizes from the length of this space. Table \ref{tab} presents such data extracted from electron microscopy and tomography of vitreous or cryo-fixed, freeze- substituted samples (also see Fig. \ref{fig:TEM}). This technique provides the most precise structural data because the  cellular structures are preserved with high fidelity \cite{Nevo2007a, Nevo2009a}. 
The dimensions of PC hexamer and APC are very similar ($\sim$2x3 and 10 nm, respectively), and allow to infer the plausible number of PC trimers forming the rod ($N_D$ in Table \ref{tab}).   We thus propose two limiting cases of possible PBS geometries, depicted in the cartoon in Fig. \ref{fig:stacking}, to fill the space between the thylakoid membranes.

\begin{figure}
\includegraphics[width=1\columnwidth]{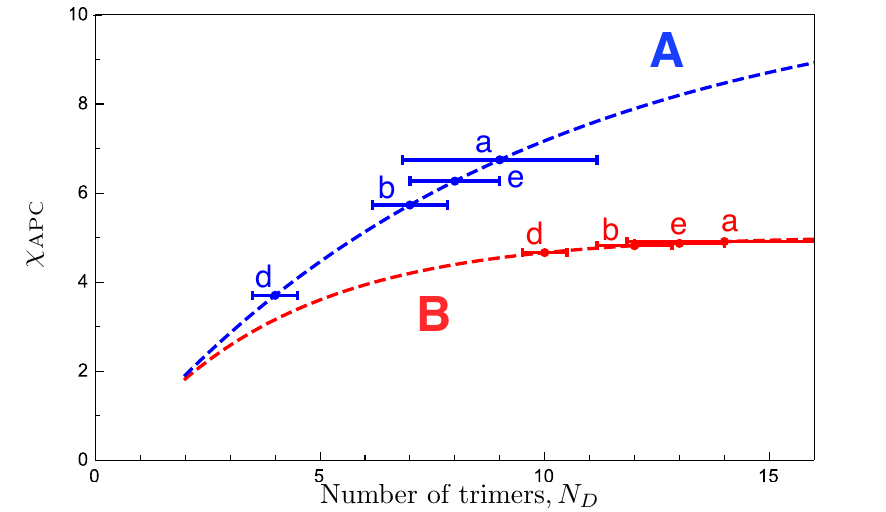}
\caption{Total excitation at the APC trap(s) as a function of the PC rod length for model A (blue, 2 traps) and model B (red, 1 trap) along with the experimental data inferred from the inter-membrane space, as presented in Table \ref{tab}. The small letters label the different species shown in Fig. \ref{fig:TEM}. \label{fig:popExp}}
\end{figure}

From a geometrical argument, stacking 3 APCs together, each modelled as a sphere of 10nm diameter,  we find that the APC core occupies about 16nm of the vertical distance. 
This value supports the available structural models, in which APC is stacked to the height of two complexes of ~16 nm (model A in Fig. \ref{fig:stacking}), to fit experiments. Indeed, this organization is supported by the measurement in the olive \textit{Synechocystis sp.} PCC6803 mutant containing only APC, and where the distance between membranes was calculated to be $\sim$32 nm  (Table \ref{tab}, and see also \cite{Liberton2013a}).  
The additional space in the wild-type species, which ranges from  $\sim$12 to  $\sim$26 nm, can be occupied by 4-9 PC trimers. 
 We can imagine a different architecture where PC rods extend all the way to the opposite membrane face, generating space for up to 14 PC hexamers in a rod (model B, Fig. \ref{fig:stacking} bottom). In reality, of course, any combination of the two extremes is possible, as is the currently used model.  
 Judging by the experimental evidence, structures with shorter PC rods are more likely. 
 Fig. \ref{fig:popExp} shows the APC excitation for the two proposed models, which can be fitted as $P_{\rm APC}(N_D) = a (1 - b \, e^{-c N_D})$, using $\{a=10.7,\, b=1.0,\, c=0.1 \}$ for model A and $\{a = 5,\, b=1.1,\, c= 0.3\} $ for model B. 
 The comparison with the measured species is only indicative, as the rates have been evaluated from the structure of cyanobacterial \textit{Thermosynechococcus vulcanus}. The stacking of model B results in an almost flat distribution and  constant APC excitation  across the presented species.  Model A yields to a larger excitation reaching the APCs, but requires the production of twice as many cores compared to model B. These results support shorter PC rods for larger excitation of the core, and is a further argument supporting the currently used model for PBS  (model C in Fig.\ref{fig:popExp}) which consists of  staggered PBS with 2-3 PCs, and is based on considerable experimental data  \cite{Olive1997a, VanDeMeene2012a}.

In conclusion, we have developed a model to study the transport efficiency in the PBS complexes as a function of the rod length. Our  kinetics analysis shows an optimal length for which the rod transport efficiency is maximal. It presents a rational to justify the evolutionary selected adaptation mechanism which consists in changing the PC:APC ratio as response to the light intensity. This model can  be extended to study organisms with much larger PBS structures composed of heterogeneous rods that contain additional PE phycobiliproteins \cite{Ong1991a, Marx2014b}, and study the PC:PE ratio as adaptation to the light colour. 
Based on the measurement of distances between the thylakoid membranes of various cyanobacterial species, we propose two  limiting stacking models of the PBS complexes within the stromal. The kinetics analysis support models with shorter rods, in line with the currently adopted PBS arrangement.   Further considerations, such as the biological cost implying such a model and the free energy of the arrangements,  are needed to discuss the most likely arrangement. There might be a trade-off between maximizing the excitation harvested and the actual cost of building super-complexes such as the  APC core, which could help further resolve the exact structure of phycobilisome complexes.

\emph{Acknowledgments.---} 
We thank I. Eisenberg and F. Caycedo-Soler for sharing the data in Ref. \cite{Eisenberg2017a} and for useful discussions. We acknowledge funding from NSF (Grant No. CHE- 1112825) and SMART IRG ID.

\bibliographystyle{apsrev4-1}

\clearpage

\setcounter{equation}{0}
\setcounter{figure}{0}
\setcounter{table}{0}
\makeatletter
\renewcommand{\theequation}{S\arabic{equation}}
\renewcommand{\thefigure}{S\arabic{figure}}
\renewcommand{\thetable}{S\arabic{table}}

\appendix

\section{Appendix A: Excitation transfer in isolated subunits}

The structure  \cite{Schirmer1986a,Schirmer1987a} and spectra of isolated chromophores \cite{Scharnagl1989a, Scharnagl1991a, Debreczeny1993a} have been used to calculate transfer rates using F\"orster theory for different assembly level, i.e. monomer, trimer and hexamer \cite{Sauer1987a, Sauer1988a, Debreczeny1995a, Debreczeny1995b}. 
\begin{figure}[b]
\includegraphics[width=1\columnwidth]{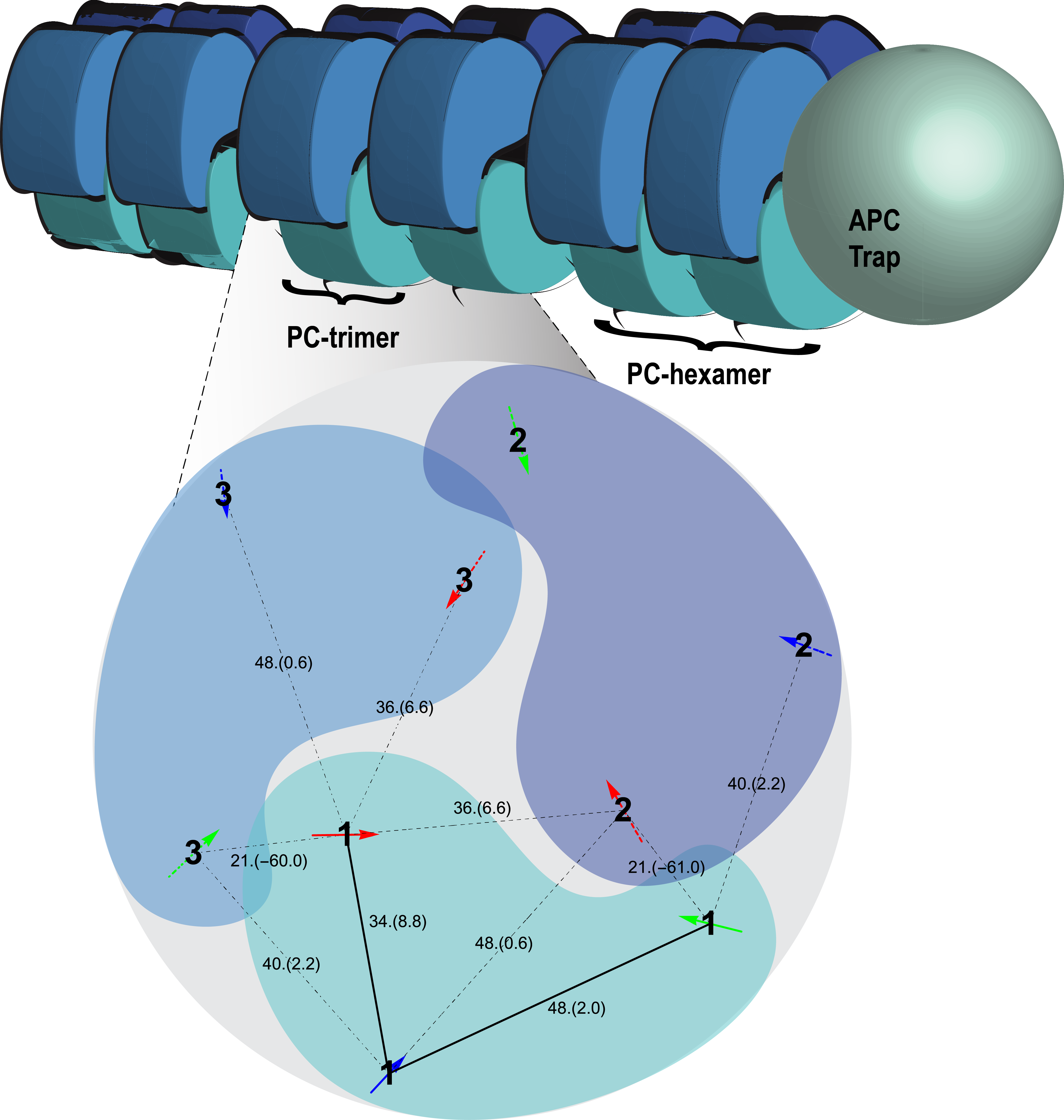}
\caption{Structure of the chromophores in the PC trimer of \textit{Thermosynechococcus vulcanus} \cite{Adir2001b} displaying the orientation of the transition dipole moments for the $\alpha_{84}$ (green), $\beta_{84}$ (red) and $\beta_{155}$ (blue) chromophores part of each $(\alpha \beta)$ monomer, numerated with bold numbers. Distances less than 50\AA \, are depicted with lines and given in \AA; the number in parenthesis are the couplings estimated from dipole-dipole interaction ($V_{dd}$ in Eq. 2, in units of cm$^{-1}$). \label{fig:structure}}
\end{figure}
The energy transfer rates within the component of the ($\alpha \beta$) monomer forming C-PC have been measured around 50-500 ps \cite{Debreczeny1993a}. 
Self-assembly into trimer result in stronger interaction between $\alpha_{84} - \beta_{84}$  of neighbouring monomeric building block within the trimer and hexamer super-structure, accelerating  the transfer rates up to the range 0,5 - 1 ps and suggesting a combination of excitation delocalization and slower incoherent F\"orster transfer  \cite{Beck1992a}. Disk-to-disk energy transfer has been modelled in structure formed by 2 or 3 disk of C-PC hexamers using stochastic computer simulation \cite{Xie2002a}, showing sub-ps quasi unidirectional transfer with an efficiency decreasing with the rod length. 
Below, we provide an analysis using the chromophore spectra derived from the aggregate conformation \cite{Eisenberg2017a}, for any arbitrary rod length, and show the length for which transport to APC is optimal.

\begin{table}
\begin{tabular}{lccc}
PC isolated from 		& $\tau_{\alpha_{84}-\beta_{84}}$ & $\tau_{\beta_{84}-\beta_{155}}$	& Refs. \\
\hline
Synechococcus 6301		& 200   $\pm20$		& 50$\pm20$	&\cite{Holzwarth1987a, Suter1987a}\\
Synechococcus Sp. PCC 7002 	& 149   $\pm2$			& 52$\pm9$	&\cite{Debreczeny1993a} \\
Mastigocladus laminosus		& 180   $\pm60$		& 57$\pm20$	&\cite{Gillbro1988a}\\
 						 & 38.9			& 20	 		&\cite{Sauer1988a}\\
Agmenellum quadruplicatum   &40.8			& 23.8		&\cite{Sauer1988a}\\
Westiellopsis prolifica		& 151				& 56			&\cite{Xia1991a}	\\
\end{tabular}
\caption{Transfer time, $\tau_{a-b}= (k_{ab} + k_{ba})^{-1}$, in units of ps for different species between the composing chromophores of the PC monomeric unit. \label{tab:isolated}  }
\end{table}

\section{Appendix B: Excitation transfer in the PC rod}

Table \ref{tab:rates} presents the major rates of transfer within the PC rod. 
\begin{table}
\begin{tabular}{rccr | r}
$k$ (ns$^{-1}$) & \multicolumn{2}{c}{Chromophores} & $k$ (ns$^{-1}$) & $(k_{ab} + k_{ba})^{-1}$ \\
\cline{2-3} 
$a\rightarrow b$ & $a$ & $b$ & $b\rightarrow a$ & (ps)\\
\hline
\hline
\multicolumn{4}{l}{Monomer $(\alpha \beta)$} \\
\hline
%
    18.34	& $^1\beta_{155}$ & $^1\beta_{84}$ &       2.28 &      48.48 \\ 
     10.85	&  $^1\alpha_{84}$ & $^1\beta_{84}$&       4.41 &      65.55 \\ 
      0.16	& $^1\alpha_{84}$ & $^1\beta_{155}$ &       0.50 &    1516.24 \\ 
       \multicolumn{4}{l}{Trimer $(\alpha \beta)_3$} \\ 
\hline
   1325.4	&  $^1\alpha_{84}$ & $^2\beta_{84} $&      538.3 &        0.5 \\ 
    1327.0	&  $^2\alpha_{84}$ & $^3\beta_{84} $&      538.9 &        0.5 \\ 
    1291.7	&  $^3\alpha_{84}$ & $^1\beta_{84} $&      524.6 &        0.6 \\ 
 \multicolumn{4}{l}{Hexamer $(\alpha \beta)_6$} \\
\hline
    1312.1	&  $^6\alpha_{84}$ & $^4\beta_{84} $&      532.9 &        0.5 \\ 
    1291.7	&  $^4\alpha_{84}$ & $^5\beta_{84} $&      524.6 &        0.6 \\ 
    1292.9	&  $^5\alpha_{84}$ & $^6\beta_{84} $&      525.1 &        0.6 \\ 
     137.7	&  $^5\alpha_{84}$ & $^2\alpha_{84}$&      137.7 &        3.6 \\ 
     137.6	&  $^3\alpha_{84}$ & $^4\alpha_{84}$&      137.6 &        3.6 \\ 
     138.2	&  $^1\alpha_{84}$ & $^6\alpha_{84}$&      138.2 &        3.6 \\ 
      34.1	&  $^1\beta_{155}$ & $^4\beta_{155}$&       34.1 &       14.7 \\ 
      34.1	&  $^5\beta_{155}$ & $^3\beta_{155}$&       34.1 &       14.7 \\ 
      33.4	&  $^2\beta_{155}$ & $^6\beta_{155}$&       33.4 &       15.0 \\ 
      13.1	&  $^1\beta_{84} $ & $^4\alpha_{84}$&       32.3 &       22.0 \\ 
      13.3	&  $^2\beta_{84} $ & $^6\alpha_{84}$&       32.6 &       21.8 \\ 
      13.1	&  $^3\beta_{84} $ & $^5\alpha_{84}$&       32.3 &       22.0 \\ 
\end{tabular}
\caption{Details of the rates between each chromophores of the PC rod. In the trimer and hexamer structures, only couples of chromophores with a rate larger than 20ns$^{-1}$ are presented here. \label{tab:rates}}
\end{table}

\section{Appendix C: Harvesting time}
We  look at the harvesting time, described from the excitation life the PC rod, and calculated with and without the APC trap, $\bar{\tau}$ and $\tau_0$, respectively. 
The mean life-time is defined as 
\begin{equation}
\bar{\tau} = \int dt \:  t \left( -\sum_{n=1}^{N_D} \dot{P}_n(t) \right) = \int dt  \left( \sum_{n=1}^{N_D} {P}_n(t) \right).
\end{equation}
Since the populations are given by the rate equation, the integral can be evaluated exactly yielding 
\begin{equation}
\bar{\tau} = \sum_{nm }(K^{-1})_{nm} P_m(0).
\end{equation}
The excitation life time with the trap, $\tau$, is obtained using $K = K^F + K^D+ K^T $; $\tau_0$ is the mean time without the APC trap, i.e. without $K^T$ in the rate matrix. The harvesting time is then readily given by \cite{Malyshev2003a}
\begin{equation}
\frac{1}{\tau} = \frac{1}{\bar{\tau} } - \frac{1}{\tau_0}.
\end{equation}

\begin{figure}
\includegraphics[width=1\columnwidth]{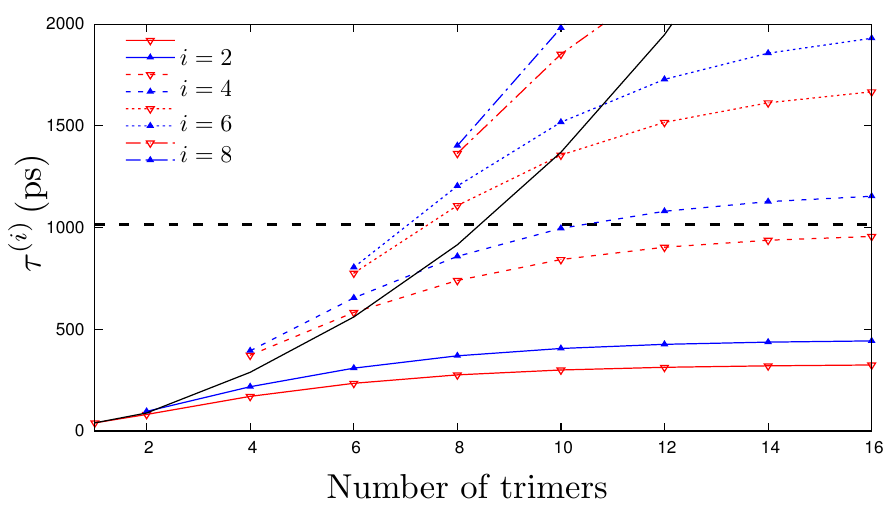}
\caption{Harvesting time $\tau^{(i)}$ by the APC trap as function of the PC rod length for different initial conditions $P^{(i)}(0)$. The excitation life time without the trap (black dashed line), $\tau_0$, is found constant and independent of the rod length; the mean harvesting time (solid black line), scales exponentially with the length of the rod.  Beyond a length of 8 trimers, only the  first 4 trimers present a harvesting time shorter that the excitation life-time in the rod, ensuring excitation transfer. \label{fig:trap} }
\end{figure}

We further label $\tau^{(i)}$ the  harvesting time after excitation of the $i-$th disk. 
In our system, the excitation life-time without the trap is independent on the length of the rod due to the relatively small transfer rate among the PC disks, often on the order of the decay rates. We find $\tau_0^{(i)} = \tau_0 = 1.0$ns, independent of the excitation condition. 
When the rod terminates with an APC trap, the excitation mean life time depends on the size of the rod and the initial condition. Figure \ref{fig:trap} presents the harvesting life time for different rod lengths, and for the excitation on each possible disk in the rod. We see that the harvesting mean time of harvesting $N$ photon on the $N$-trimer rod scales exponentially with the rod length. Beyond a length of 6 PC trimers (3 hexamers), an increase in the rod length does not  provide higher harvesting power as the harvesting time from a 4th hexamer becomes larger that the excitation life time. The optimal rod length to maximize transfer of excitation is then obtained for 3 hexamers.

\end{document}